canhep.tex
\documentstyle[11pt,epsf]{article}
\input epsf
\begin{document}
\baselineskip 100pt
\renewcommand{\baselinestretch}{1.5}
\renewcommand{\arraystretch}{0.666666666}
\parindent=0pt
{\large
\parskip.2in

\newcommand{\be}{\begin{equation}}
\newcommand{\ee}{\end{equation}}
\newcommand{\br}{\bar}
\newcommand{\fr}{\frac}
\newcommand{\lm}{\lambda}
\newcommand{\ra}{\rightarrow}
\newcommand{\al}{\alpha}
\newcommand{\bt}{\beta}
\newcommand{\pr}{\partial}
\newcommand{\hs}{\hspace{5mm}}
\newcommand{\dg}{\dagger}
\newcommand{\ve}{\varepsilon}

\hfill DTP\,97/51

\begin{center}
{\bf Soliton Solutions of the Integrable Chiral Model in 2+1
Dimensions
\footnote{ to appear in {\it Proceeding of the Soliton Conference},
Kingston 1997}}   
\end{center}

\begin{center}
T{\small HEODORA} I{\small OANNIDOU}\\
{\sl Dept of Mathematical Sciences, University of Durham,\\
Durham DH1 3LE, UK}
\end{center}

{\bf Abstract} We present soliton and soliton-antisoliton
solutions for the integrable chiral model in 2+1 dimensions with
nontrivial (elastic) scattering.
These solutions can be obtained either as the limiting cases of the ones
already constructed by Ward or by adapting Uhlenbeck's method.

\section{Introduction}

The integrable chiral model was derived by Ward \cite{W}, by dimensional
reduction from the self-dual-Yang-Mills equations in (2+2) dimensions, and
is defined as:
\be
(\eta^{\mu \nu}+\varepsilon^{\mu \nu \al}V_\al) \pr_\mu(J^{-1}\pr_\nu
J)=0,
\label{mch}
\ee
where $J$ is a map of {\bf R$^{2+1}$} into $SU(2)$, $\eta^{\mu
\nu}=\mbox{diag}(-1,1,1)$,  and  $V_\al$ is a unit vector in space-time.
In Ward's model $V_\al$ is a space-like vector, i.e. $V_\al=(0,1,0)$.
[This choice was motivated by the energy conservation.]
Notice that the standard chiral model in (2+1) dimensions has
$V_\al=(0,0,0)$.

Ward showed that the integrable chiral model has the same conserved energy
density as the standard chiral model.
This energy is given by 
\be
{\cal E}=-\fr{1}{2}\mbox{tr} \left[(J^{-1}J_t)^2+(J^{-1}J_x)^2
+(J^{-1}J_y)^2 \right].
\label{energy}
\ee
In order to ensure the finiteness of the energy of the solutions,
we require that $J \ra K+O(r^{-1})$ at spatial infinity, with $x+iy=r
e^{i \theta}$.

First of all, recall that the standard chiral model is not integrable but
Lorentz covariant.
By contrast, the existence of $V_\al$ in (\ref{mch}) breaks explicitly the
Lorentz covariance of the model but it makes it integrable;
it is associated with a linear system (see (\ref{lax})), passes  the
Painleve test, has an inverse scattering description and admits
multisoliton solutions.

One might expects the scattering of solitons of (\ref{mch}) to be
trivial, due to its integrability.
However, numerical simulation show that right-angle scattering (like the
ones observed to nonintegrable models) can occur.
This happens since the solitons in this model possess internal degrees of
freedom that determines their orientation in space.
So they can interact trivially and nontrivially depending on the
orientation of these internal parameters and on the value of the impact
parameter. 
And due to the integrability of the model these solutions should be
constructed explicitly.
From now on, by solitons we shall mean localized energy configurations
that move but we shall not imply stability of the shape or the velocity
or simple behaviour in collision.
In fact, the solutions we are going to discuss here, are localized along
the direction of motion; they are not, however, of constant seize;
their height, which is the maximum of the energy density ${\cal E}$, is
time dependent.

\section{Construction of Soliton Solutions}

Harmonic maps from ${\bf R^2}$ into Lie groups were studied by Uhlenbeck
\cite{U}, who in a seminal work, showed that all solutions of the
integrable {\it static} standard chiral model can be factorized into a
product of factors (so-called {\it $n$-unitons}), of the form
\be
J=\prod_{i=1}^{n}(1-2 R_i), 
\label{pedio}
\ee
where $R_i=(q_i^\dg \otimes q_i)/|q_i|^2$  are Hermitian
projectors, and $q_i$ are 2-dimensional vectors.
Using Lax pair she show that the projectors $R_i$ have to satisfy
first-order partial differential equations.
This can be extended to the nonstatic case (\ref{mch}).

So the 2-uniton solution of (\ref{mch}) has the form of (\ref{pedio})
with
\be 
q_1=(1,f), \hs \hs q_2=(1+|f|^2)(1,f)-2i(tf'+h)(\br{f},-1),
\ee
where $f$ and $h$ are rational meromorphic functions of $z=x+iy$.
For $f=z^p$ and $h=z^q$ (for $p>q$), the configuration consists of $(p-1)$
static solitons at the centre of mass of the system, accompanied by
$N=q-p+1$ solitons accelerating towards the ones in the middle, scattering
at an angle of $\pi/N$, and then decelerating as they separate. 
This follows from the fact that the field $J$ departs from its asymptotic
value $J_0$ when $(tf'+h) \ra 0$ which is true when either $z^{(p-1)}=0$
or $t p+z^N=0$; and this is approximately where the solitons are located.

Let us indicate how these solutions were constructed, and how others may
be obtained. 
Equation (\ref{mch}) is the consistency condition for the Lax
pair
\be
L\Psi \equiv (\lm \pr_x-\pr_y-\pr_t)\Psi=A \Psi, \hs \hs 
M\Psi \equiv (\lm \pr_t- \lm \pr_y-\pr_x)\Psi=B \Psi,
\label{lax}
\ee
where $A$ and $B$ are $2\times 2$ matrices independent of $\lm$, and the
integrability condition implies that 
$\Psi(\lm,x,y,t)|_{\lm=0}=J^{-1}(t,x,y)$.

$\bullet$ Uhlenbeck's construction, assumes that
\be
\Psi={\cal K}\,(1-\fr{2i}{\lm-i}R_n)\,...\, 
(1-\fr{2i}{\lm-i}R_2)(1-\fr{2i}{\lm-i}R_1),
\label{uni}
\ee
while the restriction for $A=(L \Psi)\Psi^{-1}$ and $B=(M \Psi)\Psi^{-1}$
to be independent of $\lm$: impose a sequence of first-order differential
equations for time dependent projector valued fields.

So, to construct time-dependent solutions of the $SU(N)$ model 
(\ref{mch}), one can start from a constant solution (0-uniton) and add to
it 1-uniton.
This solution will be nonconstant - but it is static.
Then, to this uniton, one can add a second uniton, and so on.
In \cite{I-Z} we give a 2-uniton solution of the $SU(3)$ model 
(\ref{mch}).
In fact, the number of unitons can be arbitrary.
In the static $SU(2)$ case considered by Uhlenbeck the only solutions are
those described by constant matrices (0-uniton) and factors constructed
from holomorphic functions (1-uniton).
In the model (\ref{mch}) the 0- and 1-uniton solutions are the same but
then, we
have further ones corresponding to 2- and more unitons (cf. \cite{I}).
These additional solutions are nonstatic.
We do not know at this stage, whether there is any bound on the uniton
number so that all solutions correspond to field configurations of up this
number.
 
$\bullet$ Ward \cite{W}, on the other hand, using the {\it Riemann problem
with zeroes}, proved that $\Psi$ is of the form
\be
\Psi=1-\sum_{k=1}^n (\lm-\mu_k)^{-1} \left(\sum_{l=1}^n
(\Gamma^{-1})^{kl} \br{m}^l_\al m^k_b\right), \hs
\Gamma^{kl}=(\br{\mu}_k-\mu_l)^{-1}\sum_{\al=1}^2 \br{m}^k_\al  m^l_b
\label{arx}
\ee
where $m_\al^k=(1,f_k)$ and  $f_k$ is a rational function of a complex
parameter $\omega_k$ ($\omega_k|_{\mu_k=i} \ra z$ which
correspond to a static $J$). 
These solutions pass each other without any change of direction or phase
shift.
All this assumes that the parameters $\mu_k$ are distinct and $\mu_k \neq
\br{\mu}_l$. 
The solution (\ref{uni}) is derived from (\ref{arx}) for $n=2$, by setting
$\mu_1=i+\varepsilon$, $\mu_2=i-\varepsilon$, $f_1=f+\varepsilon h$,
$f_2=f-\varepsilon h$ and by taking the limit $\varepsilon \ra 0$ (cf.
\cite{W1}, \cite{I}). 
It may seems strange that one can take the limit of a family of soliton
solutions with trivial scattering, and obtain a new one with nontrivial
scattering.
However, in \cite{I} it has been shown that as $\varepsilon \ra 0$ the
solitons disperse, shift and interact with each other.

A different and maybe more interesting problem is the existence of
soliton-antisoliton configurations for (\ref{mch}).
Roughly speaking, solitons correspond to $f$ being a function of $z$ and
antisolitons correspond to a function of $\br{z}$.
One way of proceeding is to take (\ref{arx}) with $n=2$, put
$\mu_1=i+\varepsilon$ and $\mu_2=-i-\varepsilon$  and take the limit
$\varepsilon \ra 0$ (cf. \cite{I}). 
In order $\Psi$ to be smooth on ${\bf R^{2+1}}$, it is necessary to take
$f_1=f$ and $f_2=\br{f}^{-1}-\ve h$  with $f=f(z)$ and $h=h(\br{z})$.
The corresponding field $J$ describes solitons and antisolitons
that are well separated, accelerate towards each other until they merge at
the origin and scatter at right angles as they separate again (no
radiation emission).
This is the first example in (2+1) dimensions with elastic scattering
between soliton-antisoliton solution. [Recall that for the pure ${\rm   
O}(3)$ $\sigma$-model the soliton and antisoliton after the collision 
annihilate into a wave of pure radiation.]

A topological charge  may be defined for the field $J$ by exploiting the
connection of $(\ref{mch})$ with the  ${\rm O}(3)$ $\sigma$-model
(recall that (\ref{mch}) is not a topological model).
The standard chiral model is equivalent to the ${\rm O}(4)$
$\sigma$-model, through the relation 
\be 
J=1 \,\phi_0+\mbox{\boldmath $i \sigma \cdot \phi$}.
\ee
For the ${\rm O}(4)$ model , the field at fixed time is a map
$(\phi_0,\mbox{\boldmath $\phi$}): {\bf S^3} \ra {\bf S^3}$, i.e. $\pi_2
({\bf S^3})=0$ so there is no winding number.
However, for soliton solutions that correspond to some initial embedding
of ${\rm O}(3)$ space into ${\rm O}(4)$ , there is a useful topological
quantity, as we are going to see.
For the ${\rm O}(3)$ model, the field is a map $\phi: {\bf
S^2} \ra {\bf S^2}$ and due to the homotopy relation $\pi_2 ({\bf
S^2})=0$, such maps are classified by an integer winding number, given by
\be
{\cal N}=(8\pi)^{-1}\int\epsilon_{ij}\,\mbox{\boldmath$\phi$}\cdot(\pr_i
\mbox{\boldmath $\phi$}\wedge\pr_j \mbox{\boldmath
$\phi$})\,d^2x,
\label{win}
\ee
So, if the field $J$ (at fixed time) is restricted to an ${\bf S^2}$
equator of the  ${\bf S^3}$ target space, while it never maps to
the antipodal points $\{1,-1\}$, one can define a winding number for
(\ref{mch}).
An expression for this  winding number is easy to be given, since it is
the winding number of the map after projection onto the chosen ${\bf S^2}$
equator, i.e. (\ref{win}) where $\mbox{\boldmath $\phi$} \,\ra$\,
$\mbox{\boldmath $\phi^\prime$}\equiv\mbox{\boldmath
$\phi$}/|\mbox{\boldmath $\phi$}|$.

So we see that the integrable chiral model has many interesting
solutions and these solutions can be derived using {\it analytic} methods.
These structures travel with nonconstant velocities, their size is not
constant, and they interact nontrivially (like the nonintegrable models).
The model does not posses rotational symmetry in the $x,y$ plane;
however, most explicit solutions seem to possess.
Such results might be useful for connecting integrable and nonintegrable
models, which possess soliton solutions.
In addition, they indicate the likely occurrence of new phenomena in
higher-dimensional soliton theory that are not present in 1+1 dimensions.

\section{Acknowledgements }

I am grateful to Richard Ward and Wojtek Zakrzewski for stimulating 
discussions. I also acknowledge support from EC ERBFMBICT950035.

\end{document}